# The Newsbridge - Telecom SudParis

# VoxCeleb Speaker Recognition Challenge 2022 System Description

*Yannis Tevissen*[1,2], *Jérôme Boudy*[1], *Frédéric Petitpont*[2]

[1] SAMOVAR, Telecom SudParis, Institut Polytechnique de Paris, France
[2] Newsbridge, France
`yannis.tevissen@telecom-sudparis.eu`

## Abstract

We describe the system used by our team for the VoxCeleb Speaker Recognition Challenge 2022 (VoxSRC 2022) in the speaker diarization track. Our solution was designed around a new combination of voice activity detection algorithms that uses the strengths of several systems. We introduce a novel multi stream approach with a decision protocol based on classifiers entropy.

We called this method a multi-stream voice activity detection and used it with standard baseline diarization embeddings, clustering and resegmentation. With this work, we successfully demonstrated that using a strong baseline and working only on voice activity detection, one can achieved close to state-of-the-art results.

**Index Terms**: speaker diarization, voice activity detection

## 1. Introduction

Performing speaker diarization is equivalent to answering the question of "who spoke when?" within a given audio scenario. Diarization is a key element of almost every modern speech processing application especially those which then use voice recognition algorithms.

Diarization approaches are diverse as shown in [1] and [2] but the principle is often the same. First homogeneous speech segments are being isolated within an audio file and then they are clustered to determine which segments has been pronounced by which speaker. In the past ten years tremendous improvements of diarization performances and robustness have been achieved, partially because of the development of neural networks that are now able to deal with large amount of data and especially new speech embedding features [3], [4].

Following this algorithmic pattern, front-end analysis of almost every diarization method consists in determining when there are voices within a media. This necessary step is called voice activity detection (VAD).

VAD methods are also very diverse [5]–[8] but the choice of the right algorithm for diarization pre-processing is often neglected.

The speaker diarization track 4 of VoxCeleb Speaker Recognition Challenge (VoxSRC) focuses on the analysis of online content such as talk-shows, discussions, debates, and interviews. For the challenge we decided to study the impact of the voice activity detection system on speaker diarization results.

## 2. System description

To improve diarization performances we chose to work on the preprocessing part and especially voice activity detection. We implemented a multi stream approach with a decision protocol based on classifiers entropy.
Once we had our voice activity results, we used a standard VBx system. This system contains a standard x-vectors extractor followed by a Bayesian HMM clustering. Diarization results are then resegmented to get the overlap regions right.
Our work started by testing several voice activity detection algorithms, among which pyannote 2.0 and the GPVAD systems.

### 2.1. Voice activity detection

Our work started by testing several voice activity detection algorithms, among which pyannote 2.0 and the GPVAD systems.

#### 2.1.1. Pyannote 2.0 voice activity detection

This approach is purely neural based since it relies on the PyanNet architecture described in [7]. MFCCs are used as input features for a binary classifier. This method is robust to domains mismatch and an extensive description of the recurrent neural networks is available in [7]. For this experiment, open-sourced pyannote model was used. It has been pretrained on the AMI dataset [9].

#### 2.1.2. GPVAD

To achieve a more robust voice activity detection, authors of [8] proposed a neural network approach designed with a combination of convolutional and recurrent layers and based on a weakly supervised training scheme. Trained on the Audioset dataset, this approach learns to detect 517 sound classes among which speech. The large variety of labels obtained was radically reduced to obtain a Speech/Non speech binary classifier.

#### 2.1.3. Multi-Stream voice activity detection

We noticed a complementarity between pyannote 2.0 and GPVAD voice activity detections since the first one seems to produce more constant results and the second one seems to work better for noisy audios. We decided to implement a solution to take the advantage of both VAD systems.

The two VAD systems were run in parallel. Thresholds optimized on the VoxConverse development set were applied before the decision protocol.

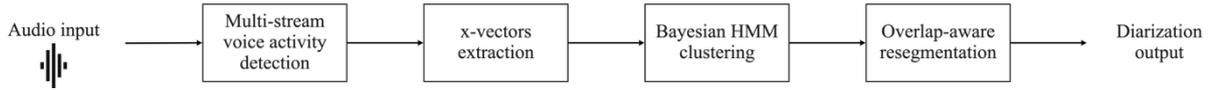

Figure 1: *System Description*

Table 1: *Thresholds used for our system. All of them were optimized on VoxConverse development set*

| Method | Onset | Offset | Min. duration on | Min. duration off |
|---|---|---|---|---|
| Pyannote VAD | 0.767 | 0.713 | - | - |
| GPVAD | 0.010 | 0.010 | - | - |
| Multi-Stream VAD | - | - | 0.182 | 0.501 |
| Resegmentation | 0.537 | 0.724 | 0.410 | 0.563 |

We introduced a decision protocol based on entropy to dynamically choose the VAD classifier. For each classifier, local entropy is computed over sliding 250ms windows. For each window, entropy $h_{k,i}$ of each binary classifier were calculated with the following formula (1).

$$h_{k,i} = -P(Speech|o_{k,i}) \log_2 P(Speech|o_{k,i}) \\ -P(Non\ Speech|o_{k,i}) \log_2 P(Non\ Speech|o_{k,i}) \quad (1)$$

Where $i$ is the time index and $k$ the classifier index of the observation denoted $o_{k,i}$.

After that, we chose for each window the classifier with the lowest entropy to obtain our multi-stream voice activity detection as depicted in Fig.3.

The two VADs were run in parallel on a Nvidia T4 GPU and an Intel Cascade Lake CPU with a maximum inference time of 14s for a single file.

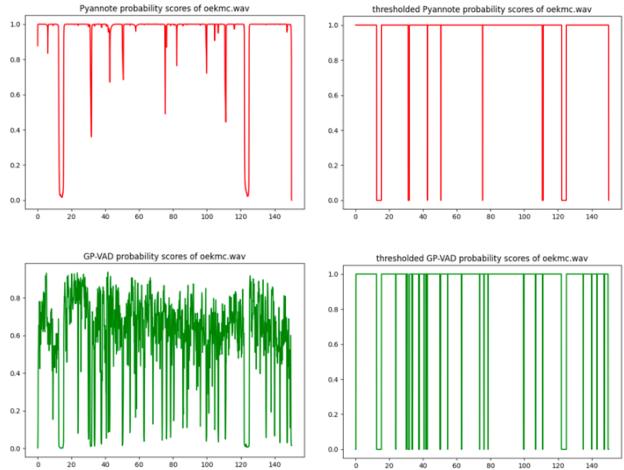

Figure 2: *Raw and thresholded voice activity detection scores from both pyannote 2.0 (in red) and GP-VAD (in green)*

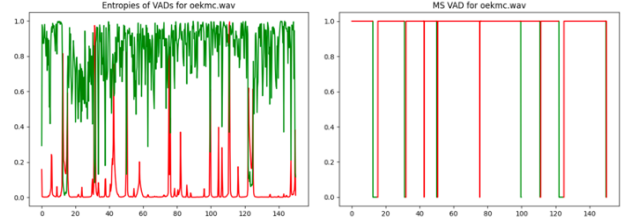

Figure 3: *Local entropies of both VAD classifiers (left) and multi-stream voice activity output (left)*

### 2.2. Speech embedding

X-vectors[4] extraction is based on 64 Mel filter banks going through a ResNet101 [10] network to create 256-values speech embeddings. The embedder is trained on VoxCeleb 1 & 2 [11]. As presented in [12], a probabilistic linear discriminant analysis (PLDA) is also applied to reduce the dimensionality before clustering.

### 2.3. Clustering

A Bayesian HMM clustering method was used as presented in [13] with one state per speaker and AHC as initialization.

### 2.4. Resegmentation

All previous components were not made to handle overlapping speech segments so we added a neural resegmentation [14] to recover these regions. It allowed us to divide by two our errors in terms of missed speech.

## 3. Results and analysis

We evaluated the proposed systems on VoxConverse 0.3 test set and report the detailed results with a 250ms collar that matches the challenge evaluation rules.

Diarization error rate (DER) and Jaccard error rate (JER) [15] are reported together with the percentages of missed speech (MS), false alarm (FA) and speaker confusion (SC).

For comparison we also report the results obtained with pyannote 2.0 neural diarization pipeline.

Table 2: *Speaker diarization results on VoxConverse test set*

| Method | MS | FA | SC | DER | JER |
|---|---|---|---|---|---|
| pyannote diarization | 2.29 | 1.31 | 4.36 | 7.96 | 43.38 |
| pyannote VAD + VBx | 3.09 | 0.79 | **2.78** | 6.66 | **29.61** |
| w/ resegmentation | **1.55** | 1.65 | 2.98 | **6.18** | 29.73 |
| GP-VAD + VBx | 3.78 | 2.30 | 3.68 | 9.76 | 31.82 |
| w/ resegmentation | **1.55** | 1.65 | 3.5& | 6.88 | 30.44 |
| **Multi-Stream VAD + VBx** | 3.05 | 1.41 | 3.09 | 7.54 | 30.22 |
| w/ resegmentation | **1.55** | 1.65 | 3.04 | 6.39 | 29.80 |

For the VoxSRC Track 4, we report the results of our three best systems.

Table 3: *Speaker diarization results obtained on the challenge platform*

| Method | DER | JER |
|---|---|---|
| Challenge baseline | 19.60 | 41.43 |
| Pyannote 2.0 VAD + VBx | 8.30 | 31.14 |
| Pyannote 2.0 VAD + VBx + resegmentation | 7.32 | 30.12 |
| Multi-Stream VAD + VBx + resegmentation | **6.62** | **29.01** |

Although all our systems achieve similar scores, the best solution for the challenge was not the best on VoxConverse test set. This can be explained by slight domain differences between the two corpuses.

With the proposed system, we achieved 6.62% on the challenge leaderboard which places it only 2% away from the best solutions. This shows that improving VAD only can lead to significant improvements of diarization scores. Moreover, we ranked 5$^{th}$ in term of JER which indicates probably a relatively fair distribution of errors across all speakers. This would need to be further investigated to determine under which conditions (noise, speaker turn duration, etc.) our multi-stream approach outperforms standard speaker diarization methods with only one VAD.

### 3.1. Reproducible research

All our experiments, systems and results are reproducible since they rely mostly on open-sourced algorithms. Thanks to their authors, VAD systems[1,2] and our baseline diarization[3] are directly available for use.

## 4. Conclusion

For this year challenge we chose to work exclusively on voice activity detection. Our work led to significant improvements of diarization scores and proved that improving diarization preprocessing was a direct way to make our system more robust. We aim to further improve this system to improve its performances on various datasets. For this, we will need to precise in which scenario, our multi-stream approach is the most relevant.

## 5. Acknowledgements

This research has received full financial support from the company Newsbridge (https://newsbridge.io/). We would also like to thank Hervé Bredin for the nice discussions and his insightful advice about our work and systems.

## 6. References

[1] X. A. Miro, S. Bozonnet, N. Evans, C. Fredouille, G. Friedland, and O. Vinyals, "Speaker Diarization: A Review of Recent Research," *IEEE Trans. Audio, Speech Lang. Process.*, vol. 20, no. 2, pp. 356–370, 2012.

[2] T. J. Park, N. Kanda, D. Dimitriadis, K. J. Han, S. Watanabe, and S. Narayanan, "A Review of Speaker Diarization: Recent Advances with Deep Learning," 2021.

[3] E. Variani, X. Lei, E. McDermott, I. L. Moreno, and J. Gonzalez-Dominguez, "Deep neural networks for small footprint text-dependent speaker verification," *ICASSP, IEEE Int. Conf. Acoust. Speech Signal Process. - Proc.*, pp. 4052–4056, 2014.

[4] D. Snyder, D. Garcia-Romero, G. Sell, D. Povey, and S. Khudanpur, "X-Vectors: Robust DNN Embeddings for Speaker Recognition," *ICASSP, IEEE Int. Conf. Acoust. Speech Signal Process. - Proc.*, vol. 2018-April, pp. 5329–5333, 2018.

[5] D. K. Freeman, G. Cosier, C. B. Southcott, and I. Boyd, "Voice activity detector for the Pan-European digital cellular mobile telephone service," *ICASSP, IEEE Int. Conf. Acoust. Speech Signal Process. - Proc.*, vol. 1, pp. 369–372, 1989.

[6] M. Lavechin, M. P. Gill, R. Bousbib, H. Bredin, and L. P. Garcia-Perera, "End-to-end domain-adversarial voice activity detection," *Proc. Annu. Conf. Int. Speech Commun. Assoc. INTERSPEECH*, vol. 2020-Octob, pp. 3685–3689, 2020.

[7] H. Bredin *et al.*, "Pyannote.Audio: Neural Building Blocks for Speaker Diarization," *ICASSP, IEEE Int. Conf. Acoust. Speech Signal Process. - Proc.*, vol. 2020-May, pp. 7124–7128, 2020.

[8] H. Dinkel, Y. Chen, M. Wu, and K. Yu, "Voice activity detection in the wild via weakly supervised sound event detection," *Proc. Annu. Conf. Int. Speech Commun. Assoc. INTERSPEECH*, 2020.

[9] I. McCowan *et al.*, "The AMI meeting corpus," 2005.

[10] K. He, X. Zhang, S. Ren, and J. Sun, "Deep residual learning for image recognition," 2016.

[11] A. Nagraniy, J. S. Chungy, and A. Zisserman, "VoxCeleb: A large-scale speaker identification dataset," *Proc. Annu. Conf. Int. Speech Commun. Assoc. INTERSPEECH*, vol. 2017-Augus, pp. 2616–2620, 2017.

[12] F. Landini *et al.*, "Analysis of the BUT Diarization System for VoxConverse Challenge," 2020.

[13] F. Landini, J. Profant, M. Diez, and L. Burget, "Bayesian HMM clustering of x-vector sequences (VBx) in speaker diarization: Theory, implementation and analysis on standard tasks," *Comput. Speech Lang.*, vol. 71, pp. 1–39, 2021.

[14] L. Bullock, H. Bredin, and L. P. Garcia-Perera, "Overlap-Aware Diarization: Resegmentation Using Neural End-to-End Overlapped Speech Detection," *ICASSP, IEEE Int. Conf. Acoust. Speech Signal Process. - Proc.*, vol. 2020-May, pp. 7114–7118, 2020.

[15] O. Galibert, "Methodologies for the evaluation of Speaker Diarization and Automatic Speech Recognition in the presence of overlapping speech," *Proc. Annu. Conf. Int. Speech Commun. Assoc. INTERSPEECH*, no. August, pp. 1131–1134, 2013.

---

[1] https://github.com/pyannote/pyannote-audio
[2] https://github.com/RicherMans/GPV
[3] https://github.com/BUTSpeechFIT/VBx